\documentclass{article}

\usepackage[utf8]{inputenc}
\usepackage[pdftex]{graphicx}
\usepackage{citesort}
\usepackage{multirow} 
\usepackage{rotating} 
\usepackage[all]{xy}
\usepackage{amsmath} 
\usepackage{amssymb}
\usepackage{setspace}
\usepackage[rightcaption]{sidecap}
\usepackage{bibspacing}

\begin{document}

\title{
  White Paper: The Universal Recommender \\ 
  \large{A Recommender System for Semantic Networks}
} 

\author{
  Jérôme Kunegis, Alan Said and Winfried Umbrath \\ 
  \small{DAI-Labor, Technische Universität Berlin, Germany} \\
  \small{\texttt{\{jerome.kunegis,alan.said,winfried.umbrath\}@dai-labor.de}}
}

\date{September 2009}

\maketitle

\begin{abstract} 
We describe the Universal Recommender, a recommender system for semantic
datasets that generalizes domain-specific recommenders such as
content-based, collaborative, social, bibliographic, lexicographic,
hybrid and other recommenders.  In contrast to existing recommender
systems, the Universal Recommender applies to any dataset that allows a
semantic representation.  We describe the scalable three-stage
architecture of the Universal Recommender and its application to
Internet Protocol Television (IPTV).  To achieve good recommendation
accuracy, several novel machine learning and optimization problems are
identified.  We finally give a brief argument supporting the need for
machine learning recommenders. 
\end{abstract}

\section{Introduction} 
In the field of information retrieval, a recommender system is defined
as a system that is able to find entities in a dataset that may be of
interest to the user~\cite{b201}.  In contrast to search engines,
recommender systems do not base their results on a \emph{query}, instead
they rely on implicit and explicit connections between users and items,
such as ratings or other past interactions.  Research and development in the
area of recommender systems has grown in recent years, as
witnessed by the creation of a high-profile conference devoted to them.

In the general case, a recommender system applies to a dataset described
by a data model containing entities (such as users and items) and
relationships (such as ratings and social links).  In the simplest
recommender system, data consists of one relationship type
connecting one or two entity types.  In more complex cases, the dataset
contains multiple relationship types connecting any number of entity
types.

The simple case, with only one relationship type, corresponds to several
well-studied recommendation subproblems, such as link prediction,
collaborative filtering, citation analysis, etc.  In the case of
multiple relationship types, hybrid recommenders are normally used.  As
we will show, most hybrid recommenders however are specific to one data
model and cannot be generalized to other data models.  Therefore, each
time a new data model is introduced, a new hybrid recommender has to be
developed.

To avoid these problems we propose the \textbf{Universal
Recommender}~(UR), a recommendation engine with the following
features: 
\begin{itemize} 
  \item It applies to datasets with any number of entity and
    relationship types.   
  \item It learns all its parameters without human interaction.   
\end{itemize}
The first feature ensures the recommender can be applied to future
datasets whose structures are unknown.  The second feature is more
critical; since a recommender using semantic datasets has a large set of
parameters, a complex recommender runs the risk of either overfitting or
having very low recommendation quality.  The main challenges of the
Universal Recommender are therefore a series of machine learning
problems related to the complexity of semantic networks, and a series of
optimization problems that must be solved to make the recommender
scalable.  We give a brief justification for machine learning
recommenders by interpreting non-learning recommenders as a source of
additional data, enhancing machine learning recommenders.

We begin by reviewing typical recommendation datasets and
tasks, and give known solutions to specific recommendation problems.  We
then describe the unified semantic data model and the architecture
of the Universal Recommender.  We continue by
identifying the machine learning and optimization
problems it will have to solve.  Finally we describe the case study
of the Internet Protocol Television (IPTV) recommender system.

\section{State-of-the-art Recommenders} 
In this section, we review classical recommendation settings
motivating the complexity of typical recommendation datasets. We also
review existing solutions for recommendation problems that apply to
specific dataset types.

We describe all cases in the context of the Internet Protocol Television
(IPTV) recommender system~\cite{b430}.  The IPTV systems delivers
television programs over the Internet instead of traditional broadcast
signals.  Using the flexibility of the Internet Protocol, IPTV
devices have the possibility to provide additional services on top of
TV viewing.  One of these additional services are
recommender systems.  In our supposed setting, the goal is thus
to recommend a TV program to a user.  In the following examples we
describe classical recommendation settings applied to our IPTV scenario.

\subsection{Content-based Filtering} 
The first kind of recommender we describe uses the content of items to
generate recommendations, similarly to search engines.  While search
engines require the user to enter a specific keyword for searching,
content-based recommenders usually take keywords from another source,
for instance a user profile containing words describing user interests,
or from items already seen or rated.

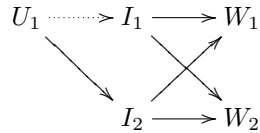
\begin{SCfigure}[1.5][h!]
  $
    \xymatrix{ 
      U_1 \ar@{.>}[r] \ar[rd] & I_1 \ar[r] \ar[rd] & W_1 \\ 
      & I_2 \ar[ru] \ar[r] & W_2 
    } 
  $
  \caption{The entity-relationship diagram of a dataset used by
    content-based recommenders. }
  \label{fig:er:content-based}
\end{SCfigure}

Figure~\ref{fig:er:content-based} shows a situation in which a user--item
recommendation is found by comparing common features of two items.  In
this entity-relationship (ER) diagram, arrows represent known relationships and
the dotted arrow represents the relationship to predict.  $U$, $I$ and
$W$ represent users, items and words respectively.

Content-based filtering has traditionally been applied to document
recommendation, using the tf--idf measure as edge weights.  In our IPTV 
system, each TV program has a description of its content.  This
example nevertheless shows a limitation of the content-based approach for
IPTV; due to the fact that descriptions are much shorter than typical
documents, the tf--idf measure will be less accurate.

\subsection{Collaborative Filtering} 
\begin{SCfigure}[2.4][h!]
  $
    \xymatrix{
      U_1 \ar@{.>}[r] \ar[rd] & I_1 \\ 
      U_2 \ar[ru] \ar[r] & I_2 
    } 
  $
  \caption{The entity-relationship diagram of a dataset used by
    collaborative recommenders. }
  \label{fig:er:collaborative}
\end{SCfigure}
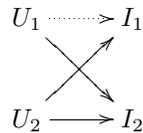

While the content-based approach is simple (essentially being a search
engine), only making use of one user's relations to items,
collaborative filters attempt to make use of all known
relations between users and items.  For instance, if our IPTV system
tracks the programs watched by each user, this information can be used
directly giving collaborative filters, as shown in
Figure~\ref{fig:er:collaborative}.

The general idea is the following: If we know which TV programs
user~$U_1$ has watched, we can use this information directly, without
content information, to make connections between items.  To do this, a
collaborative recommender system must consider the behavior of other
users.  If user~$U_2$ has seen the same TV program~$I_2$ as user~$U_1$,
then we can recommend other TV programs seen by $U_2$.  The
resulting recommender system does not need any content information.
Therefore, collaborative recommenders are often used in scenarios where
little or no content is available, movies or jokes for
instance~\cite{b7}.  For our IPTV recommender, this means we do not have
to rely on content descriptions, and can thus recommend TV programs
lacking a description.

Furthermore, a collaborative recommender can make use of 
ratings.  Compared to the typical \emph{has-seen} information,
ratings have the advantage of also admitting negative values, modeling
dislike.  

\subsection{Social Networks and Link Prediction} 
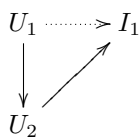
\begin{SCfigure}[2][h!]
  $
    \xymatrix{ 
      U_1 \ar@{.>}[r] \ar[d] & I_1 \\ 
      U_2 \ar[ur]
    } 
  $
  \caption{The entity-relationship diagram of a dataset used by social
    recommenders. }
  \label{fig:er:social}
\end{SCfigure}

Another type of recommender is given when links are known between users.
For instance, if IPTV users maintain a buddy list, we can recommend
the favorite items of a given user's buddies.  This type of
recommendation is particularly useful when trust is important.  In this
case, a trust measure can be defined between users denoting the level
of confidence a user has in another user.
Methods to compute trust include local measures~\cite{b325,b367} and
global approaches, which often generalize the PageRank
measure~\cite{b235,b236,b237}.  
Figure~\ref{fig:er:social}
gives the associated entity-relationship diagram. 

In addition to friendship and trust, which are positive relationships,
we may allow users to mark other users as foes (or enemies),
representing distrust~\cite{b325,kunegis:slashdot-zoo}.


\subsection{Lexicographic Information} 
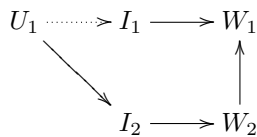
\begin{SCfigure}[1.5][h!]
  $
  \xymatrix{ 
    U_1 \ar@{.>}[r] \ar[dr] & I_1 \ar[r] & W_1 \\ 
    & I_2 \ar[r] & W_2 \ar[u] 
  } 
  $
  \caption{The entity-relationship diagram of a dataset used by
    lexicographic recommenders. }
  \label{fig:er:lexicographic}
\end{SCfigure}

While words contained in descriptions may be used to find similar TV
programs, the words themselves may be modeled as interlinked entities:
Some words are synonyms, antonyms, etc.~\cite{b437} These relationships
may be used to enhance a content-based recommender by recommending
items of a related topic using different glossaries~\cite{b436,b441}.
Figure~\ref{fig:er:lexicographic} gives the associated
entity-relationship diagram, in which $W$ denotes words. 

We might even go as far as mapping words in different languages to each
other, using information from a dictionary.  This would allow the IPTV
system to recommend programs in other languages. 

\subsection{Hybrid Recommenders} 
\begin{SCfigure}[1.5][h!]
  $
  \xymatrix{ 
    U_1 \ar@{.>}[r] \ar[rd] \ar[d] & I_1 \ar[r] & W_1 \\ 
    U_2 \ar[ru] \ar[r] & I_2 \ar[ru] \ar[r] & W_2 \ar[u]
  }
  $
  \caption{The entity-relationship diagram of a dataset used by hybrid
    recommenders. }
  \label{fig:er:hybrid}
\end{SCfigure}
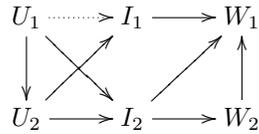

In many recommender systems, several of
the previously described dataset types are known.  For instance, a
recommender system may have user ratings for items and at the same
time content information about items.  Recommenders that apply to such
datasets are called hybrid recommenders.  While hybrid recommenders exist for
many combinations of entity and relationship types
(e.g.~\cite{b335,b30,b379,wetzker2009a,b222}), none of these can be
applied to all semantic networks since they are not generic.  One of
many possible 
data models is shown in Figure~\ref{fig:er:hybrid}. 

\subsection{Semantic Networks} 
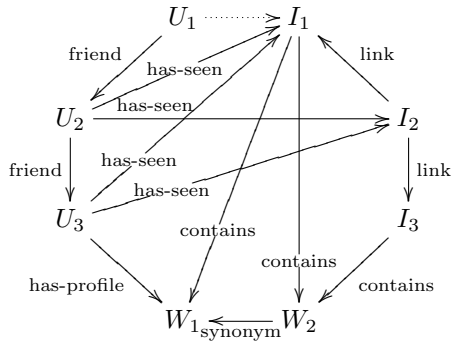
\begin{SCfigure}[.7][h!]
  $
  \xymatrix{ 
    & U_1 \ar@{.>}[r] \ar[ld]_{\text{friend}} & I_1
    \ar[lddd]|(.7){\text{contains}} \ar[ddd]|(.8){\text{contains}} \\ 
    U_2 \ar[rru]|{\text{has-seen}} \ar[rrr]^(.25){\text{has-seen}}
    \ar[d]_{\text{friend}} & & & I_2 \ar[lu]_{\text{link}}
    \ar[d]^{\text{link}} \\ 
    U_3 \ar[rruu]|(.3){\text{has-seen}} \ar[rrru]|(.3){\text{has-seen}}
    \ar[rd]_{\text{has-profile}} & & & I_3 \ar[ld]^{\text{contains}} \\ 
    & W_1 & W_2 \ar[l]^{\text{synonym}}
  } 
  $
  \caption{A semantic network consisting of entities and relationships
    of different types. }
  \label{fig:er:semantic}
\end{SCfigure}

In the general case, datasets can be modeled as a set of entities
connected by relationships.  While in simple datasets relationships are
similar (e.g. the \emph{has-seen} relationship), more complex networks
almost always contain multiple relationship types.  This is especially
true when several datasets are combined.  The result is a semantic
network, where multiple relationship types connect multiple entity
types.  An example is shown in Figure~\ref{fig:er:semantic}.

Several ongoing projects collect data from various sources and integrate
them to semantic networks, DBpedia~\cite{b332}, YAGO~\cite{b333} and
Freebase~\cite{b427} for instance.  Probabilistic models that
apply to semantic networks exist~\cite{b391,b438}, they have however not been
used for recommendation.

Semantic networks are general enough to represent the datasets described
previously in this section.  Therefore, the Universal Recommender will
use a semantic representation of datasets.  The following section describes
this representation in detail. 

\section{Unified Semantic Representation of Datasets} 
In order to write a recommender system that supports all the use cases
described in the previous section, we define a unified semantic
representation of the datasets the Universal Recommender will be applied
to.

\begin{itemize} 
  \item Datasets consist of entities and relationships, each
    connecting two or more entities.  
  \item Entities are grouped into multiple entity types $E_i$.  
  \item Relationships are grouped into multiple relationship types
    $\mathcal R_i$, each connecting a predefined number of fixed entity
    types.   
  \item Relationships may be symmetric or asymmetric,
    corresponding to undirected and directed relations.  
  \item Relationships may be unweighted or weighted, and weights may be
    negative.   
  \item Entities and
    relationships may both be annotated with attributes, for instance
    timestamps of ratings or the age of users.  
\end{itemize}

Note that this definition not only includes binary relationships, but
also higher-order relationships (e.g. tag assignments between users,
tags and items.)  Table~\ref{tab:reltypes} gives some examples of
relationship types between users, items and words.
Table~\ref{tab:format-weight} gives examples of relationship types by
the number of different entity types they connect (unipartite,
bipartite) and the range of edge weights.  Table~\ref{tab:examples}
gives an overview of traditional data mining applications that can be
interpreted as special cases of recommendation.

\begin{table}
  \centering \small 
  \caption{ 
    Common relationship types in recommender systems, arranged by the
    entity types they connect.  Only unipartite and bipartite relationship
    types are shown.  Unipartite relationship types are on the diagonal
    entries of the table.
  } 
  \begin{tabular}{|c||l|l|l|}
    \hline & \textbf{User} & \textbf{Item} & \textbf{Word} \\ \hline \hline
    \multirow{4}{*}{\textbf{User}} & Social network & & \\ & Trust network &
    & \\ & Email network & & \\ & Profile ratings & & \\ \hline
    \multirow{4}{*}{\textbf{Item}} & Explicit feedback & Citations & \\ &
    Implicit feedback & Hyperlinks & \\ & Authorship & & \\ & Commercial
    selling data & & \\ \hline \multirow{2}{*}{\textbf{Word}} & Search
    history & tf--idf & WordNet \\ & & Categories & \\ \hline
 \end{tabular} 
 \label{tab:reltypes} 
\end{table}

\begin{table} 
  \centering \small 
  \caption{ 
    Common relationship types by
    the number of entity types they connect and the range of admitted edge
    weights.  
  } 
  \begin{tabular}{|c||l|l|l|} 
    \hline &
    \multicolumn{3}{|c|}{\textbf{Number of connected entities}} \\ \hline &
    \multicolumn{2}{|c|}{\textbf 2} & \multicolumn{1}{|c|}{\textbf 3} \\
    \hline & \multicolumn{1}{|c|}{\textbf{Unipartite}} &
    \multicolumn{1}{|c|}{\textbf{Bipartite}} &
    \multicolumn{1}{|c|}{\textbf{Tripartite}} \\ \hline \hline
    \multirow{2}{*}{\textbf{Unweighted}} & Friendship & Authorship &
    Folksonomy \\ & & Categories & \\ \hline
    \multirow{2}{*}{\textbf{Weighted}} & Profile rating & Rating & \\ &
    Trust levels & & \\ \hline \multirow{2}{*}{\textbf{Positive}} &
    Communication & View history & Clickthrough data \\ & & & \\ \hline
    \multirow{2}{*}{\textbf{Signed}} & Friend/foe network & Like/dislike &
    Contextual rating \\ & Trust/distrust & & \\
    \hline 
  \end{tabular} 
  \label{tab:format-weight} 
\end{table}

\begin{sidewaystable}
  \centering 
  \small 
  \caption{ 
    Compilation of common relationship types, the entity types they
    connect, and the traditional data mining applications using only that
    relationship type as a dataset.
  } 
  \begin{tabular}{|p{.18\textwidth}|p{.18\textwidth}|p{.18\textwidth}|p{.18\textwidth}|p{.18\textwidth}|}
    \hline 
    \textbf{Dataset} & \textbf{Relationship types} & \textbf{Entity
      types} & \textbf{Weight range} & \textbf{Application} \\ 
    \hline \hline
    Explicit feedback & Ratings & Users, items & Signed &
    Collaborative filtering \\ 
    \hline
    Implicit feedback & View, save, etc.
    & Users, items & Unweighted & Collaborative filtering, recommendation \\
    \hline 
    Features & tf--idf & Text documents, phrases & Positive & Document
    classification \\ 
    \hline
    Social network & Friendship & Users & Unweighted & Link prediction,
    community building \\
    \hline
    Signed social network & Friendship, enmity & Users & $\{-1, +1\}$ &
    Link sign prediction, community building \\
    \hline
    Web & Hyperlinks & Web pages & Unweighted & Link prediction,
    ranking \\ 
    \hline
    Citation network & References & Scientific
    publications & Unweighted & Bibliographic analysis \\ 
    \hline
    Trust network & Trust, distrust & Users & Unweighted or $\{-1,+1\}$
    & Trust measures \\ 
    \hline
    Interaction network & Communication (e.g. email) &
    Users & Positive integer & Link prediction, network analysis \\ 
    \hline
    User profile ratings & Ratings of user profiles & Users & Signed &
    Expert finding, date site recommendation \\ 
    \hline
    Collaboration graph &
    Authorship & Authors, publications & Unweighted & Community building \\
    \hline 
    Search history & Searches & Users, queries, phrases & Unweighted
    & Query reforming, query recommendation \\ 
    \hline
    Taxonomy & Category
    membership & Items, categories & Unweighted & Classification \\ 
    \hline
    Lexical network & Synonymy, antonymy, etc. & Words & Unweighted & Query
    reforming, lexical search \\
    \hline
    Folksonomy & Tag assignment & Users,
    items, tags & Unweighted & Tag and item recommendation, trend prediction \\ 
    \hline
  \end{tabular} 
  \label{tab:examples} 
\end{sidewaystable}

\section{Architecture}
\begin{figure}[h!]
  \begin{displaymath}
    \xymatrix{
      *+[F]{\text{Semantic dataset}} \ar@/^/[d]^{\text{decomposition}} 
      & \mathcal{R}_i \in \mathbb{R}^{E_1^i \times E_2^i} 
      \\
      *+[F]{\text{Recommender model}} \ar@/^/[d]^{\text{clustering}} 
      & U_i, V_i \in \mathbb{R}^{E_i \times k}; \Sigma
      \\
      *+[F]{\text{Recommender index}} \ar@/^/[d]^{\text{recommender}} 
      & \text{Hierarchy of entities}
      \\
      *+[F]{\text{Recommendation}}
      & (e_1, w_1); (e_2, w_2); \ldots
    }
  \end{displaymath}
  \caption{
    Computational flow diagram of the Universal Recommender.  First the
    dataset is decomposed into a recommender model.  In this model,
    entities are clustered giving a recommender index.  Finally, a
    recommender computes recommendations using the recommender index.
  }
  \label{fig:flow}
\end{figure}
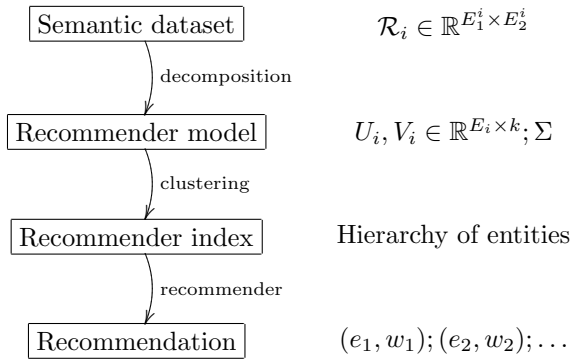

Based on the dataset structure described in the last section, the
Universal Recommender is built on a scalable three-stage architecture, 
shown in Figure~\ref{fig:flow}. 

The basis of any recommender system is a dataset, in our case it is a
semantic one.  The final output of the Universal Recommender are
recommendations.  In the first stage, the dataset is mapped to a
recommender model, which can be interpreted as a decomposition in the
general sense.  This model is then used to build a recommender index,
which allows recommendations to be computed quickly in the third stage.
The next sections describe these various steps in detail.

\section{Universal Latent Decomposition}
In this section, we describe the general decomposition approach for
semantic networks used by the Universal Recommender.  The idea consists
in representing entities in a latent space, in which relationships are
predicted by using the scalar product.  In other words, if we associate
a vector of length $k$ to every entity, we compute a prediction between
two entities using the scalar product.  This approach has two
consequences:

\begin{itemize} 
  \item Computation of the latent model can
    be interpreted as a decomposition of the adjacency matrix of the
    complete network, allowing us to use known graph kernels.  
  \item The predictions made by the latent model have to be mapped to
    recommendations.  This is described in
    Section~\ref{sec:optimization}.  
\end{itemize}

The following state-of-the-art recommendation algorithms
can be described as the decomposition of a dataset into a latent model:  

\begin{itemize} 
  \item The singular value decomposition (SVD) and eigenvalue
    decomposition (EVD)~\cite{b179} and their applications to principal
    component analysis (PCA) and latent semantic indexing (LSI)~\cite{b433}.   
  \item Graph kernels such as the exponential kernel, the von Neumann
    kernel, path counting and rank reduction methods~\cite{b263,b289,b306,b413}.  These 
    can be applied to the eigenvalue or singular value decomposition of
    graphs, and their parameters can be learned
    efficiently~\cite{kunegis:spectral-transformation}.    
  \item Methods based on the Laplacian matrix such as
    the commute time and resistance
    distance~\cite{b105,b283,kunegis:netflix-srd}, the heat diffusion kernel~\cite{b137}
    and the random forest kernel~\cite{b190}.  
  \item Probabilistic approaches such as probabilistic latent semantic
    analysis (PLSA)~\cite{b347}, and latent Dirichlet
    allocation (LDA)~\cite{b358}.   
  \item Other matrix decompositions such as nonnegative matrix
    factorization~\cite{b393}, maximum margin matrix
    factorization~\cite{b197} and low-rank approximations with missing
    values~\cite{b178}.   
  \item Higher-order decompositions 
    such as parallel factor analysis (PARAFAC)~\cite{b435}, the Tucker
    decomposition~\cite{b434} and others~\cite{b293}.  
\end{itemize}

\subsection{Example}
In the example of the IPTV recommender, we give a derivation of a
recommendation algorithm using the singular value decomposition.  Let
$U$ be the user set, $I$ the item set and $W$ the set of words.  Then
the dataset is given by the following adjacency matrices: the ratings
$\mathcal{R} \in \mathbb{R}^{U\times I}$, the buddies $\mathcal{B} \in
\{0,1\}^{U \times U}$ and the features $\mathcal{F} \in \{0,1\}^{I\times
W}$.  The weighted matrix $\mathcal R$ is then normalized to
$\bar{\mathcal R}$ and aggregated with the other adjacency matrices into
a single matrix $A \in \mathbb R ^{(U+I+W)\times(U+I+W)}$.

\begin{displaymath} A = \begin{pmatrix} w_B \mathcal B & w_R
\bar{\mathcal R} & \\ w_R \bar{\mathcal R}^T & & w_F \mathcal F \\ & w_F
\mathcal F^T & \end{pmatrix} \end{displaymath} where $w_X > 0$ is the
weighting of relationship type $X$.

This matrix is then decomposed, giving latent vectors for all three
entity types.  The approximation or decomposition used may be any of
those described above.  For simplicity, we adopt the notation of the
singular value decomposition.

\begin{displaymath} A = U \Sigma V^T = \begin{pmatrix} U_U \\ U_I \\
U_W \end{pmatrix} \Sigma \begin{pmatrix} V_U \\ V_I \\
V_W \end{pmatrix}^T \end{displaymath}

$U_X$ and $V_X$ are latent vectors of dimension $X\times k$, where $k$
is the number of latent dimensions computed.  These vectors can then be
used for computing recommendations.  To compute a rating prediction for
the user--item pair $(u,i)$, we would use $U_U(u) \cdot V_I(i)$.
Relationship types that connect more than two entity types have to be
reduced from hypergraphs to graphs in this model.  Possible reductions
are the star and clique expansions~\cite{b283}. 

\section{The Machine Learning Approach}
Here we describe the machine learning problems associated with the
Universal Recommender. 
While in unirelational networks a matrix
decomposition approach is a common procedure to recommendation, its
application to semantic networks raises additional issues: 

\begin{itemize} 
  \item Weights and sparsity patterns of different relationship types
    may be different, in which case each relationship type has to be
    \textbf{normalized} separately.  
  \item Since edge weights of different relationship types are usually not
    comparable, the question of finding the \textbf{relative weights} $w_X$
    arises.  
\end{itemize}

To motivate the machine learning approach to recommenders, consider the
case of a ``dumb'' recommender with hardcoded recommendations.  The
administrator of a recommender system may be tempted to implement
hardcoded recommendations, thinking that such a recommender may be more
pertinent than a learning recommender.  However we now have an
additional problem: How will the administrator choose the hardcoded
recommendations?  In practice he will choose the preferences of himself
or another user, i.e. enter the items \emph{someone} thinks are good.
But then the question becomes: Why would other users have the same taste
as this one user?  In fact, users do not all have the same taste and
effectively, finding which users have similar tastes amounts to writing
a collaborative recommender.  Therefore, the hardcoded recommendations
are not necessarily useful as recommendations for every user, but can be
used indirectly by a collaborative algorithm to provide better
recommendations for everyone.  In other words, trying to hardcode
recommendations in one part of the recommender will make machine
learning algorithms more useful in other parts of the system,
underlining the importance of the following machine learning problems.  

\subsection{Learning Normalizations}
In recommender systems that apply to unirelational datasets with edge
weights such as ratings, a common first step consists in additive
normalization.  Given edge weights $a_{ij}$, additive normalization
computes new edge weights $b_{ij} = a_{ij} - \tilde a_{ij}$, where
$\tilde a_{ij}$ is a simple approximation to $a_{ij}$, e.g. a row or
column mean.

In most recommenders, this step is usually kept simple, such as
subtracting the overall rating mean.  In semantic networks, each weighted
relationship type may need separate normalization, and the overall
normalization problem becomes non-trivial as the number of parameters
increases with the number of relationship types.

\subsection{Learning Relative Weights} 
In unirelational datasets, all edges have the same semantics, and an
algebraic or probabilistic decomposition algorithm can use this fact to
compute a low-rank model of the data.  In semantic networks however, such
an algorithm would implicitly assume that edges have the same
semantics, which in practice only works if the different relationship
types have a similar weight range and degree distribution.

In order to apply these algorithms to semantic networks, the different
relationship types have to be weighted separately.  The weights $w_X$
depend on the characteristics of the subnetwork (e.g. the degree
distribution), but also on overall considerations, such as whether a
particular relationship type is useful for recommendation.  Different
weights must also be applied to different relationship types connecting
the same entity types.

These weights can be hardcoded using domain-specific knowledge.  For
instance in the IPTV case, by knowing that ratings contribute more to
recommendations than the \emph{has-seen} relationship type.  These
assumptions are however difficult to justify purely from expert
knowledge.  To validate these assumptions, we would have to evaluate
recommenders that use varying values of these parameters.  If we do
this, we automatically learn which parameter values are best, and can
discard the expert knowledge.  We therefore propose the Universal
Recommender to learn relative weights automatically, in order to avoid
being dependent on domain-specific knowledge, and to validate
domain-specific knowledge if present.

Examples of different relationship types connecting users and items are
\emph{has-seen}, \emph{has-recorded} and \emph{has-bookmarked}.  While a
human IPTV expert could set these relative weights by hand, learning the
weights is a worthwhile machine learning problem in itself.

\section{Optimization Problems} 
\label{sec:optimization}
In addition to the machine learning problems which assure that
recommendations actually correspond to user expectations, the following
optimization problems must be solved to ensure the scalability of the
Universal Recommender.

\begin{itemize}
\item The computation of the latent recommender model must be
  asynchronous.  In other words, updates to the data model must be
  incorporated into the recommender model without recomputation of the
  whole recommender model.  In practice, the recommender model is built
  iteratively, and the data model is read at each iteration, ensuring
  that changes are incorporated into the recommender immediately.
\item While rating predictions can be computed in constant time for a
  given recommender model (using the scalar product), computations of
  recommendations are more complex.  The underlying problem consists of
  finding a vector maximizing the scalar product with a given vector.
  This problem is similar but not identical to the metric nearest-neighbor
  problem.  A common approach consists in clustering the set of entities.
\end{itemize}

The following subsections describe possible solutions to these
optimization problems. 

\subsection{Iterative Update of the Recommender Model} 
To compute recommendations in a dataset, a recommender model is
built from the dataset.  This model building step may be slow, but the
resulting model can be used to compute any number of recommendations rapidly.  If the
dataset changes, for instance when users rate additional items, the
model would have to be recomputed.

To avoid this overhead, we propose a recommender model that can be
updated iteratively.  In fact, most matrix decomposition and low-rank
approximation problems can be solved iteratively, giving a recommender
model where updates arise naturally from the decomposition
algorithm~\cite{b135,b178,b347}.

In this context, the role of the recommender model is analogous to
PageRank for search engines~\cite{b133}.  The PageRank is a vector of
entities (web pages) that can be updated by iterative algorithms
(i.e. power iteration).  In the case of the Universal Recommender, the
model consists of a set of $k$ vectors corresponding to the latent
spaces of the rank reduction method, and updates can be performed in a
way consistent with the underlying algorithm.

\subsection{Recommender Index} 
The goal of a recommender is to compute recommendations.  Functionally,
a recommender takes an entity as input (a user) and outputs a list of
ranked entities (items).  
While rating prediction has received
attention in itself (e.g. in the Netflix Prize), they are only useful to a
recommendation system insofar as they can be used to rank items.  To
find the top $k$ items that a user would rate with high scores, all $n$
items have to be considered.  Since runtime of recommendation should not
depend on $n$, a \emph{recommender index} has to be used.

A recommender index must solve the following problem: Given $n$ vectors
$a_i$ and a vector $x$, find the top $k$ vectors $a_i$ such that $x\cdot
a_i$ is maximal.  A similar problem with the scalar product replaced by
the Euclidean distance is known as \emph{nearest neighbor search}.

We conjecture that this problem can be solved analogously to the nearest
neighbor problem by partitioning the unit
hypersphere into regions containing a constant number of vectors $a_i$ and 
storing, for each region, the list of adjacent regions in a way that
requires only linear memory in the number of regions and dimensions. 

\section{Case Study: IPTV} 
\begin{figure}
  \centering
  \includegraphics[width=.9\textwidth]{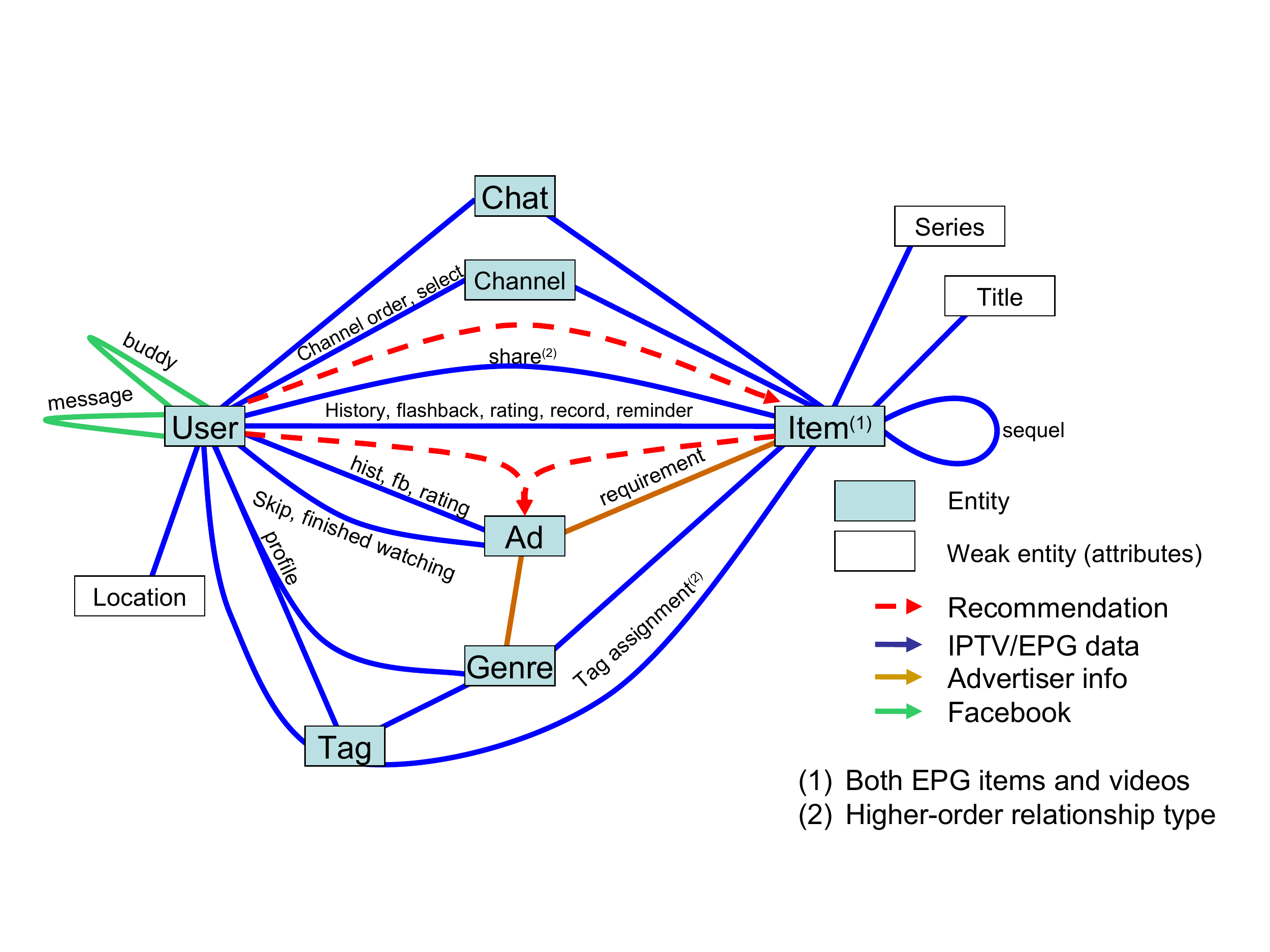}
  \caption{
    Entity-relationship diagram of the IPTV system.  Entities are
    represented by nodes and relationships by edges.
    Recommendation use cases are drawn as dashed red arrows.  Any dataset
    usable by the Universal Recommender can be represented by such a
    graph.
  }
  \label{fig:iptv-er}
\end{figure}

In this section we describe the IPTV recommender system as an example
setting for the Universal Recommender.
In the Internet Protocol Television system, users can watch TV programs
over the Internet.  In addition to the functionality provided by regular
television, our IPTV includes a semantic \emph{recommender
  system} based on the Universal Recommender.  Figure~\ref{fig:iptv-er}
shows the entities and relationships present in the IPTV system, along
with the main recommendation scenario.

\begin{figure}
  \centering 
  \includegraphics[width=.8\textwidth]{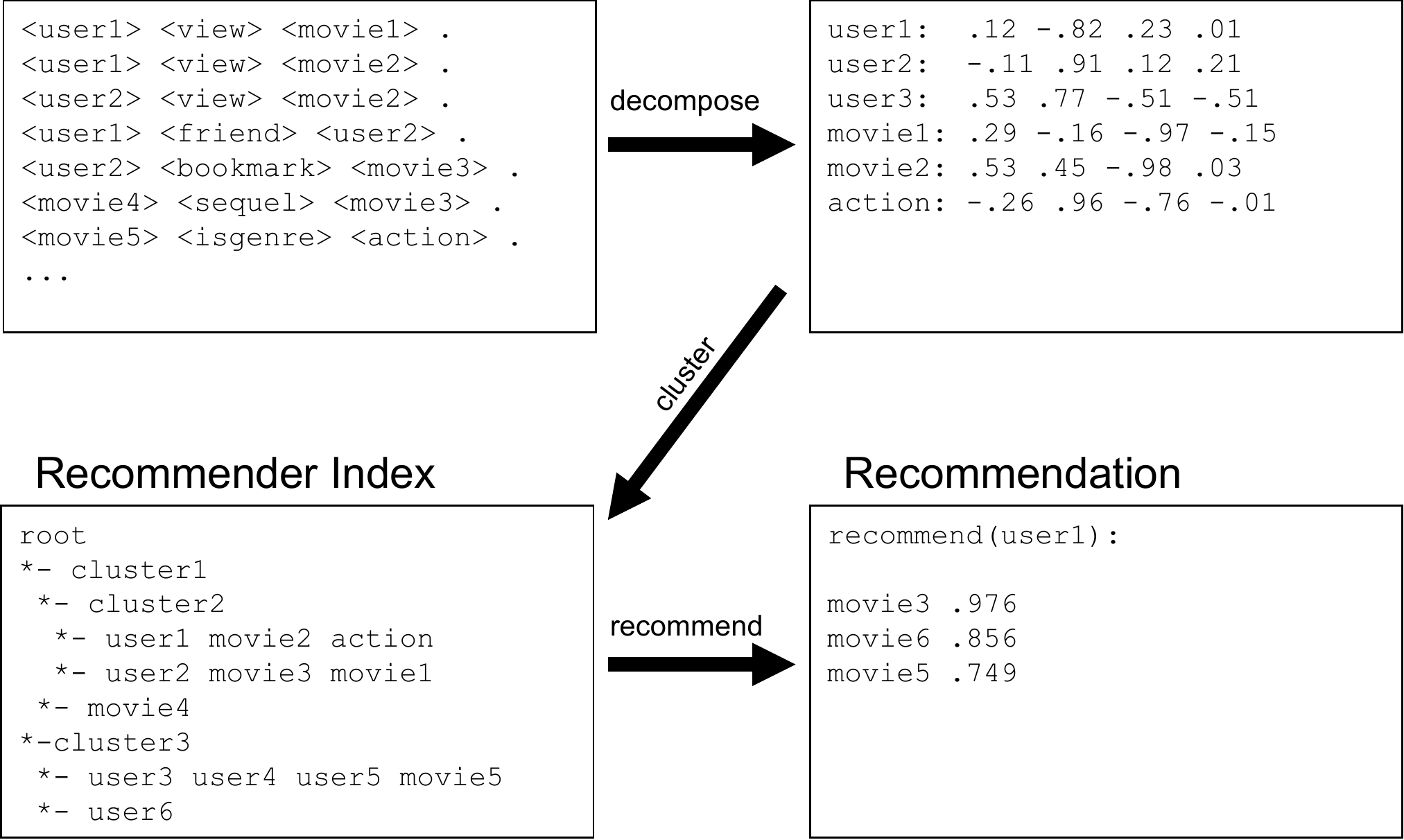}
  \caption{
    Example data flow in the IPTV recommender. From top left to bottom
    right:  
    The dataset contains users, movies and genres, connected by 
    \emph{view}, \emph{friend}, \emph{bookmark}, \emph{sequel}
    and \emph{isgenre} relationships. 
    The entities of the semantic dataset are mapped into a
    latent space of three dimensions.  The entities are then clustered
    hierarchically into a recommender index.  The recommender index is
    then used to compute recommendations for \texttt{user1}. 
  }
  \label{fig:overview}
\end{figure}

This example shows characteristics found in many recommender system
datasets: The primary entity types are users and items, which are TV
programs in this case.  The main relationship types connect users and
items.  In our example these are \emph{view}, \emph{flashback},
\emph{rating}, \emph{record} and \emph{reminder} events.  This scenario
shows a common feature of recommender systems: several relationships
connect the same entity types.  Other relationship types connect
secondary entities such as \emph{location}, \emph{genre}, \emph{series}
and \emph{title}.  User--user relationships are represented by message
events and buddy lists, both common in recommender systems.  This
dataset also contains higher-order relationship types, in the form of
tag assignments and shared events.  Recommendations in this dataset can
be computed by using a recommender model and a recommender index, see
Figure~\ref{fig:overview} for an example.

This example also shows how difficult it is, in general, to find and
build a good hybrid recommender system out of simple recommender
systems, because 
the number of relationship types is too large to be optimized by trial
and error.

\section{Conclusion} 
By describing the Universal Recommender, we hope to make clear the need
for a generic recommender system that applies to semantic datasets.  The
Internet Protocol Television example shows that datasets available in
recommender systems are usually complex and require not only hybrid
recommenders, but generic recommenders that apply to any dataset.  As we
have seen, many state-of-the-art recommenders appear as special cases of
our proposed Universal Recommender.

To implement the Universal Recommender, a unified representation of
datasets is needed, which we propose to be semantic.  As many
recommendation algorithms are based on the notion of embedding entities
in a low-dimensional space, we adopt a latent representation for the
Universal Recommender that covers these recommendation algorithms.

Several hard machine learning and optimization problems have to be
solved to implement the Universal Recommender successfully.  We showed
how machine learning recommenders arise in the case of trying to
hand-optimize a recommender, as a hardcoded recommendation algorithm can
be interpreted as part of the underlying dataset, enhancing machine
learning recommenders for the general recommendation problem.  We thus
come to the conclusion that the Universal Recommender will be able to
match and eventually exceed the performance of dataset-specific
recommenders, if these problems are solved. 

\begin{spacing}{0.97}
\bibliographystyle{acm} 
\bibliography{kunegis}
\end{spacing}

\end{document}